\documentclass{emulateapj}

\slugcomment{ApJ, in press}
\usepackage{apjfonts}

\shorttitle{Fragmentation of eccentric accretion disks}
\shortauthors{Alexander et al.}

\begin{document}

\title{Self-gravitating fragmentation of eccentric accretion disks}

\author{Richard~D.~Alexander\altaffilmark{1}, Philip~J.~Armitage\altaffilmark{2}, Jorge Cuadra, and Mitchell~C.~Begelman\altaffilmark{2}}
\affil{JILA, 440 UCB, University of Colorado, Boulder, CO 80309-0440}
\altaffiltext{1}{Leiden Observatory, Universiteit Leiden, Niels Bohrweg 2, 2300 RA, Leiden, the Netherlands}
\altaffiltext{2}{Department of Astrophysical and Planetary Sciences, University of Colorado, Boulder, CO 80309-0391}
\email{rda@strw.leidenuniv.nl}

%%%%%%%%%%%%%%%%%%%%%%%%%%%%

\begin{abstract}
We consider the effects of eccentricity on the fragmentation of gravitationally unstable accretion disks, using numerical hydrodynamics.  We find that eccentricity does not affect the overall stability of the disk against fragmentation, but significantly alters the manner in which such fragments accrete gas.  Variable tidal forces around an eccentric orbit slow the accretion process, and suppress the formation of weakly-bound clumps.  The ``stellar'' mass function  resulting from the fragmentation of an eccentric disk is found to have a significantly higher characteristic mass than that from a corresponding circular disk.  We discuss our results in terms of the disk(s) of massive stars at $\simeq0.1$pc from the Galactic Center, and find that the fragmentation of an eccentric accretion disk, due to gravitational instability, is a viable mechanism for the formation of these systems.
\end{abstract}


%%%%%%%%%%%%%%%%%%%%%%%%%%%%

\section{Introduction}\label{sec:intro}
The relative proximity of the Galactic Center (henceforth GC) provides a unique opportunity for ``close-up'' study of the processes that influence the formation of galaxies and black holes.  Recent advances in telescope technology have enabled the resolution of individual stars in the crowded GC environment, and these new data have produced several puzzles \citep[e.g.,][]{ghez98,genzel03,ghez05,paumard06}.  In particular, the detection of large numbers of young, massive stars is strongly suggestive of recent star formation activity at or close to the GC, despite the fact that the extreme physical conditions at the GC (temperature, tidal shear, etc.) prohibit the formation of stars by the same mechanisms that form stars in the solar neighborhood.  The data show that a small number (few tens) of B-type stars exist very close to the GC (within $\simeq0.01$pc; these are the so-called ``S-stars''), and a larger population ($>100$) of massive O, B and Wolf-Rayet (WR) stars exist at slightly larger radii ($\simeq0.1$pc).  Many of this second population of stars are known to form a coherent ring or disk \citep[e.g.,][]{genzel03} and, while the existence of a second such disk of stars remains controversial \citep[e.g.,][]{paumard06,lu06}, the existence of at least one stellar disk seems secure.

A popular theory posits that the stellar disk(s) formed via fragmentation of an accretion disk due to gravitational instability \citep[e.g.,][]{ps77,lb03,sb87,sb89,gt04,nayak06,levin07}.  However, we have recently demonstrated that the observed eccentricities of the stellar orbits \citep[typically $\gtrsim 0.3$,][]{bel06} are inconsistent with dynamical relaxation from a circular initial configuration \citep{aba07}, unless the stellar mass function (MF) is extremely top-heavy, much more so than is inferred from observations \citep{ns05,paumard06}.  We suggested that fragmentation of an eccentric accretion disk could explain this discrepancy, and here seek to investigate the effects of eccentricity on disk fragmentation.

In recent years gravitational instabilities in accretion disks have been the subject of much theoretical and numerical research, primarily in the context of protoplanetary disks and planet formation \citep[see, e.g., the review by][]{durisen_ppv}.  Some details remain contested, but the basic physical processes are now well understood.  In order for a gravitationally unstable disk to fragment, two conditions must be satisfied.  Firstly, the disk must be sufficiently massive and/or cold that the \citet{toomre64} criterion is satisfied:
\begin{equation}
Q = \frac{c_{\mathrm s} \Omega}{\pi G \Sigma} \lesssim 1
\end{equation}
Here $c_{\mathrm s}$ is the local sound speed, $\Omega$ the local orbital frequency, and $\Sigma$ the disk surface density.  This criterion ensures that the disk is gravitationally unstable, but in order for the instability to lead to fragmentation the disk must also cool rapidly, or else compressional heating will stabilize the disk against fragmentation.  \citet{gammie01} used a local analysis to show that the cooling timescale of the disk must satisfy
\begin{equation}
t_{\mathrm {cool}} \lesssim 3 \Omega^{-1} \, ,
\end{equation}
and subsequent work using global simulations has verified the validity of this criterion \citep{rice03,lr04,lr05}.  The criterion also depends weakly on the adopted equation of state \citep{gammie01,rice05}, and studies of protoplanetary disks have shown that this dependence can be non-trivial in real disks \citep{boley07}.  In the case of an eccentric disk, the circularization energy provides an additional, and potentially important, source of heating, and this remains largely unstudied to date.

Theoretical models of disk fragmentation around black holes typically consider circular disks with near-equilibrium initial conditions, implicitly assuming that such a configuration results from some sort of ``accretion event'' \citep[such as the capture of a molecular cloud, e.g.,][]{nayak06,nayak07,levin07}.  In the case of the GC, the so-called circumnuclear ring at a few pc from Sgr A$^*$ is thought to be the likely gas reservoir for such accretion events \citep[][]{morris93,sanders98}, but the dynamics of the infall process is not well understood.  Moreover, some models of cloud infall predict the formation of coherent, eccentric disks around the central black hole \citep[e.g.,][]{sanders98}.  We expect such disks to be gravitationally unstable, and star formation via fragmentation of an eccentric disk would provide an elegant solution to the problem of the stellar orbits discussed in \citet{aba07}, although it may leave some issues, such as the $z$-distribution of the orbits, unresolved.  In this paper, therefore, we seek to understand the dynamics of gravitational instability and fragmentation in eccentric disks.  In Section \ref{sec:sims} we describe our numerical simulations, and present the results in Section \ref{sec:res}.  We discuss the implications of our results for models of star formation at the GC, as well as the limitations of our analysis, in Section \ref{sec:dis}, and summarize our conclusions in Section \ref{sec:summary}.

%%%%%%%%%%%%%%%%%%%%%%%%%%%%
\section{Simulations}\label{sec:sims}
Our simulations make use of the publicly-available smoothed-particle hydrodynamics (SPH) code {\sc Gadget2} \citep{springel05}.  The code has been modified to allow the use of a simple cooling prescription (as outlined below), which is valid only in a disk geometry.   We adopt the standard \citet{mg83} prescription for the SPH viscosity (with $\alpha_{\mathrm {SPH}}=1.0$), using the ``Balsara-switch'' \citep{balsara95} to limit the artificial shear viscosity (as specified in Equations 11--12 of \citealt{springel05}), and adopt sufficiently high numerical resolution that the transport of angular momentum due to numerical dissipation is much less than that expected from self-gravitating angular momentum transport \citep{lr04,lr05,nelson06}.  As demanded by \citet{nelson06}, we allow for a variable gravitational softening length, and fix the SPH smoothing and gravitational softening lengths to be equal throughout.  We use the standard Barnes-Hut formalism to compute the gravitational force tree \citep[as described in][]{springel05}, and use $N_{\mathrm {ngb}}=64\pm2$ as the number of SPH neighbours.  Lastly, in order to avoid close orbits around the ``black hole'' limiting the time-step to an unreasonably small value, we use a single sink particle as the central gravitating mass.  This particle simply accretes all gas particles that pass within its sink radius  \citep[as described in][]{cuadra06}, and is merely a numerical convenience that has no physical effect on the simulation.  We set the sink radius of the ``black hole'' to be 1/4 of the inner disk radius (or semi-major axis, if $e\ne0$), which has a value of 1 in our scale-free simulations (see Section \ref{sec:sim_dets}).  In practice fewer than 0.1\% of the particles are swallowed by the sink particle in any of our simulations.  The simulations were run on the {\sc tungsten} Xeon Linux cluster at NCSA\footnote{See {\tt http://www.ncsa.uiuc.edu/}}, using 64 parallel CPUs for the highest-resolution runs.

\subsection{Initial conditions}\label{sec:ics}
We set up our initial disk as follows, using an approach that closely mirrors that of \citet{rice05}.  We adopt a surface density profile
\begin{equation}
\Sigma(a) \propto a^{-1}
\end{equation}
and a disk temperature that scales as 
\begin{equation}
T(a) \propto a^{-1/2} \, ,
\end{equation}
where $a$ is the orbital semi-major axis.  The disk sound speed $c_s \propto T^{1/2}$ is normalized so that the Toomre parameter
\begin{equation}
Q = \frac{c_{\mathrm s} \Omega}{\pi G \Sigma}
\end{equation}
is equal to 2.0 at the outer boundary.  (Here $\Omega$ is the orbital frequency.)  Using this prescription, the Toomre parameter scales approximately as
\begin{equation}
Q \propto a^{-3/4}
\end{equation}
in the initial disk.  With this set-up the initial disk is marginally gravitationally stable, and cools into instability with time.

In order to set up such a disk in SPH, we first define a disk mass $M_{\mathrm d}$, and set the mass of each gas particle to be $m = M_{\mathrm d}/N_{\mathrm {SPH}}$, where $N_{\mathrm {SPH}}$ is the number of gas particles.  We then divide the disk into concentric annuli, and distribute the particles in the annuli according to the surface density profile.  Within each annulus, particles are distributed such that the mass flux around the annulus is constant.  For an orbit of given semi-major axis $a$, eccentricity $e$, and azimuthal angle $\phi$, the radius is given by
\begin{equation}
r(\phi) = \frac{1-e^2}{1+e\cos\phi} a \,
\end{equation}
and the radial and azimuthal components of the velocity are given by
\begin{equation}
v_r = e \sin \phi \sqrt{\frac{G M_{\mathrm {enc}}}{r(1+e\cos\phi)}}\, ,
\end{equation}
\begin{equation}
v_{\phi} = \sqrt{\frac{G M_{\mathrm {enc}} (1+e\cos\phi)}{r}} 
\end{equation}
respectively, where $M_{\mathrm {enc}}$ is the mass enclosed by the orbit (the mass of the central object plus the mass of the disk at semi-major axis $<a$).  Note that, as our disks are sufficiently massive to be self-gravitating, the disk mass can cause a significant perturbation to the Keplerian potential of the central mass.  (Strictly these expressions only apply to a spherically symmetric mass distribution, but they are sufficiently accurate for our purposes.)  The particles are then distributed vertically by randomly sampling a hydrostatic equilibrium (Gaussian) density distribution with scale-height $H = c_{\mathrm s}/\Omega(r)$ (although we note that this is not strictly an equilibrium configuration when $e\ne0$).

\subsection{Cooling}\label{sec:cooling}
In order to draw comparisons with previous studies, we adopt a simple cooling prescription of the form
\begin{equation}
\left(\frac{du}{dt}\right)_{\mathrm {cool},i} = -\frac{u_i}{t_{\mathrm {cool}}} \, ,
\end{equation}
where $u_i$ is the internal energy of particle $i$, and the cooling timescale $t_{\mathrm {cool}}$ is given by
\begin{equation}
t_{\mathrm {cool}} = \frac{\beta}{\Omega} \, .
\end{equation}
Here the constant $\beta$ is an input parameter: previous studies have found that the fragmentation boundary in circular disks lies at $\beta \simeq 3$, with a weak dependence on the equation of state adopted \citep{gammie01,rice03,rice05}.

In the case of an eccentric disk, however, there is an ambiguity in this definition of the cooling timescale.  One can choose either to use $\Omega(r) = \sqrt{G M_{\mathrm {bh}}/r^3}$, or $\Omega(a) = \sqrt{G M_{\mathrm {bh}}/a^3}$.  The first case results in a cooling timescale which varies around the orbit, while in the second case $t_{\mathrm {cool}}$ is constant for individual orbital streamlines.  It is not entirely clear which of these is most appropriate when compared to a real disk, as this depends on the coolants involved.  However, it seems unlikely that the rate at which an individual fluid element cools will vary on a timescale shorter than the dynamical timescale, so we adopt the second prescription for most of our modelling.  We do, however, run a model with varying cooling time as a test case.  

Consequently, we define the cooling timescale as
\begin{equation}
t_{\mathrm {cool}} = \frac{\beta}{\Omega(a)} = \beta\sqrt{\frac{a^3}{G M_{\mathrm {bh}}}}\, ,
\end{equation}
and compute the semi-major axis $a$ of a given particle directly from the instantaneous orbital elements (based on the assumption of Keplerian orbits).  We adopt an adiabatic equation of state throughout, with adiabatic index $\gamma=5/3$.

\subsection{Simulation details}\label{sec:sim_dets}
\begin{table}[t]
\centering
  \begin{tabular}{|lccccccc|}
  \hline
Simulation & $N_{\mathrm {SPH}}$ & $e$ & $\beta$ & $a_{\mathrm{max}}$ & $M_{\mathrm d}/M_{\mathrm {bh}}$ & Min.~smoothing & $N_{\mathrm {runs}}$ \\
 & & & & & & length? & \\

  \hline
{\sc circ} & 500,000 & 0 & 3 & 25.0 & 0.2 & No & 3 \\
{\sc circb5} & 500,000 & 0 & 5 & 25.0 & 0.2 & No & 1 \\
{\sc ecc.25} & 500,000 & 0.25 & 3 & 25.0 & 0.2 & No & 1 \\
{\sc ecc.50} & 500,000 & 0.5 & 3 & 25.0 & 0.2 & No & 3 \\
{\sc ecc.50b5} & 500,000 & 0.5 & 5 & 25.0 & 0.2 & No & 1 \\
{\sc ecc.50var} & 500,000 & 0.5 & 3 & 25.0 & 0.2 & No & 1 \\
{\sc ecc.75} & 500,000 & 0.75 & 3 & 25.0 & 0.2 & No & 1 \\
{\sc circhr} & $5\times10^6$ & 0 & 3 & 5.0 & 0.05 & Yes & 1 \\
{\sc ecchr} & $5\times10^6$ & 0.5 & 3 & 5.0 & 0.05 & Yes & 1 \\

\hline
\end{tabular}
\caption{{\rm List of simulations run, showing resolution and physical parameters for each simulation.  $N_{\mathrm {SPH}}$ is the number of SPH particles used to model the gas disk, and $N_{\mathrm {runs}}$ is the number of different random realizations of the initial conditions that were used in each case.  The simulation {\sc ecc.50var} used the cooling prescription that results in a cooling time which varies around the eccentric orbit; all the others used a cooling time that was constant along the orbital streamlines.}}
\label{tab:sims}
\end{table}
In order to study the effects of eccentricity on disk fragmentation, we have conducted a number of simulations.  The parameters of these simulations are specified in Table \ref{tab:sims}, and are summarized below.  The simulations are scale-free: we adopt a system of units where the length unit is the semi-major axis of the inner disk edge, the mass unit is that of the central black hole, and the time unit is the orbital period at $a=1$.

Firstly, we conducted a large suite of simulations at moderate resolution, using 500,000 gas particles.  In order to resolve the fragmentation process properly we adopted a rather massive (and therefore thick) disk, with $M_{\mathrm d}/M_{\mathrm {bh}}=0.2$, and simply stopped the calculations when the density of collapsed regions became high enough to restrict the timestep to an unfeasibly small value.  We adopted a dynamic range of 25 in semi-major axis for these simulations.  For a circular disk ($e=0$) we conducted three identical simulations with $\beta=3$ (i.e.,~that are marginally unstable to fragmentation), each using a different random realization of the initial distribution of gas particles (i.e.,,~a different noise field), and similarly performed three such simulations with $e=0.5$.   In addition, we performed a single simulation with $e=0.25$, another with $e=0.75$, and a further simulation with $e=0.5$ using the cooling prescription that varies around the disk orbit (see Section \ref{sec:cooling}).  Lastly, we performed two simulations with a longer cooling time of $\beta=5$ \citep[which should be stable against fragmentation in the circular case;][]{rice03}: one with $e=0$, and one with $e=0.5$.  

As seen in Section \ref{sec:res_low} below, these low-resolution simulations allow us to infer a great deal about the effects of eccentricity on disk fragmentation, but do not run for long enough, or produce enough fragments, to allow measurement of a MF.  Consequently, we also ran two further simulations at much higher resolution, using $5\times10^6$ gas particles and a somewhat smaller radial range (with the outer edge of the disk at 5 times the semi-major axis of the inner edge).  We adopted the lightest disk that could be resolved safely throughout (at a factor of $\simeq2$ better than the \citealt{nelson06} criterion for spatial resolution in the vertical direction), and so adopted $M_{\mathrm d}/M_{\mathrm {bh}}=0.05$.  If we scale this to the GC system, this gives an SPH particle mass of 0.03M$_{\odot}$ (assuming a black hole mass of $3\times10^6$M$_{\odot}$).  We note that this is slightly more massive than the disks adopted by previous such simulations \citep[e.g.,][who adopt $M_{\mathrm d}/M_{\mathrm {bh}}=0.01$]{nayak07}, and also slightly more massive than current observational estimates of the total stellar mass in the GC disk(s) \citep[which range from $\simeq0.003$--0.03$M_{\mathrm {bh}}$, e.g.,][]{ndcg06,paumard06}.  However, such observations may under-estimate the mass of the initial gas disk as, given the age of the stellar disk(s) \citep[$\simeq 6$Myr,][]{paumard06}, it is reasonable to assume that the most massive stars that formed are no longer present.  Moreover, given the scale-free nature of our simulations, it is unlikely that this small over-estimate of the disk mass will have a significant effect on our results.

Additionally, in order to follow the fragmentation process beyond the initial collapse phase in these high-resolution simulations, we imposed a minimum SPH smoothing length of 0.001.  This is well below the Hill radius expected for even the least massive clumps, and merely acts to limit collapse to densities that would stop the calculation; the clumps essentially have density profiles which are ``flat-topped'' within this radius.  We prefer this to a sink particle approach because bound material can become unbound from fragments as they move around an eccentric orbit (due to the variable shear), and sink particles of the type used by, for example, \citet{nayak07} cannot lose mass in this manner (by construction).  We conducted two of these high-resolution simulations: one of a circular disk, and one with $e=0.5$.

%%%%%%%%%%%%%%%%%%%%%%%%%%%%
\section{Results}\label{sec:res}
\subsection{Code Tests}
As a test of our numerical method, we first set out to re-produce previously obtained results.  In the circular case, we find that the simulation with $\beta=5$ was indeed stable against fragmentation, and instead produced quasi-steady angular momentum transport via spiral density waves.  This simulation was allowed to run for 5 {\it outer} orbital periods (i.e.,~625 inner orbital periods) and showed no evidence of fragmentation, suggesting that this configuration is indeed stable against fragmentation over long timescales.  By contrast, all of the simulations with $\beta=3$ were unstable to fragmentation, typically breaking into fragments after around 90 inner orbital periods\footnote{The exact time at which fragmentation occurs is somewhat arbitrary, as it merely reflects the time required for the initial configuration to cool into instability (the disk initially has $Q\simeq 20$ in the inner regions).  However, the fact that different random realizations of the initial conditions result in fragmentation at the same time supports our assertion that the simulations are numerically converged.}.  Reflections of density waves from the inner edge of the disk act to stabilize the region within a few $H$ of the boundary, so that the first fragments form at radii of $\simeq3$--5 (where the orbital timescale is approximately 10 times that at the inner boundary).  Thus we find that our numerical procedure was successful in reproducing previous results, namely that the fragmentation boundary for a gravitationally unstable disk with $\gamma=5/3$ lies at $t_{\mathrm {cool}}\Omega \simeq 3$ \citep[e.g.,][]{gammie01,rice03}.

\subsection{Low-resolution runs}\label{sec:res_low}
The low resolution runs make use of a rather massive (and therefore thick) disk and, if they fragment at all, are not expected to produce very many fragments (as the most unstable length scale is always of order $H$).  Consequently, the analysis of these low resolution runs is limited to if, when and where they fragment, and the number of fragments produced: there are too few fragments to allow detailed analysis of their properties.  In these runs the density is allowed to increase without limit, so the simulations simply stop when the density becomes sufficiently high to force a very small timestep.  Typically this occurs 1--2 (local) orbital periods after the first fragments form, and only a small fraction of the gas is accreted into clumps before the simulations stop.

In order to compare the fragmentation properties of runs with different eccentricities, we require a simple method to quantify the behavior (masses, velocities, positions, etc.) of the fragments that form.  We define fragments (or ``clumps'') to be bound objects containing at least 128 SPH particles (i.e.,~with at least double the number of nearest neighbors).  The clumps are first identified by locating peaks in the density distribution, and are then tested for bound-ness.  For simplicity we calculate the potential energy of pairs of particles using a slightly simplified gravitational potential of the form $Gm/(r+h)$, where $h$ is the local SPH smoothing length: this is sufficiently accurate for our purposes, and is faster to compute than convolution with the rather complex smoothing function used by {\sc Gadget2}.  The masses of individual clumps are determined by ranking particles by total energy (potential plus kinetic plus thermal) and iterating outwards until the first unbound particle is reached.  We define the position and velocity of each clump as the mean values of all of the SPH particles bound to that clump.  While somewhat cumbersome, we prefer this method of identifying clumps to a geometric one (using, for example, concentric shells) as it allows accurate treatment of non-spherical clumps, and we apply this method identically to all of our simulations (both here and in the high-resolution runs discussed below).

\subsubsection{Fragmentation boundary for $e\ne0$}
\begin{figure}
\includegraphics[angle=270,width=\hsize]{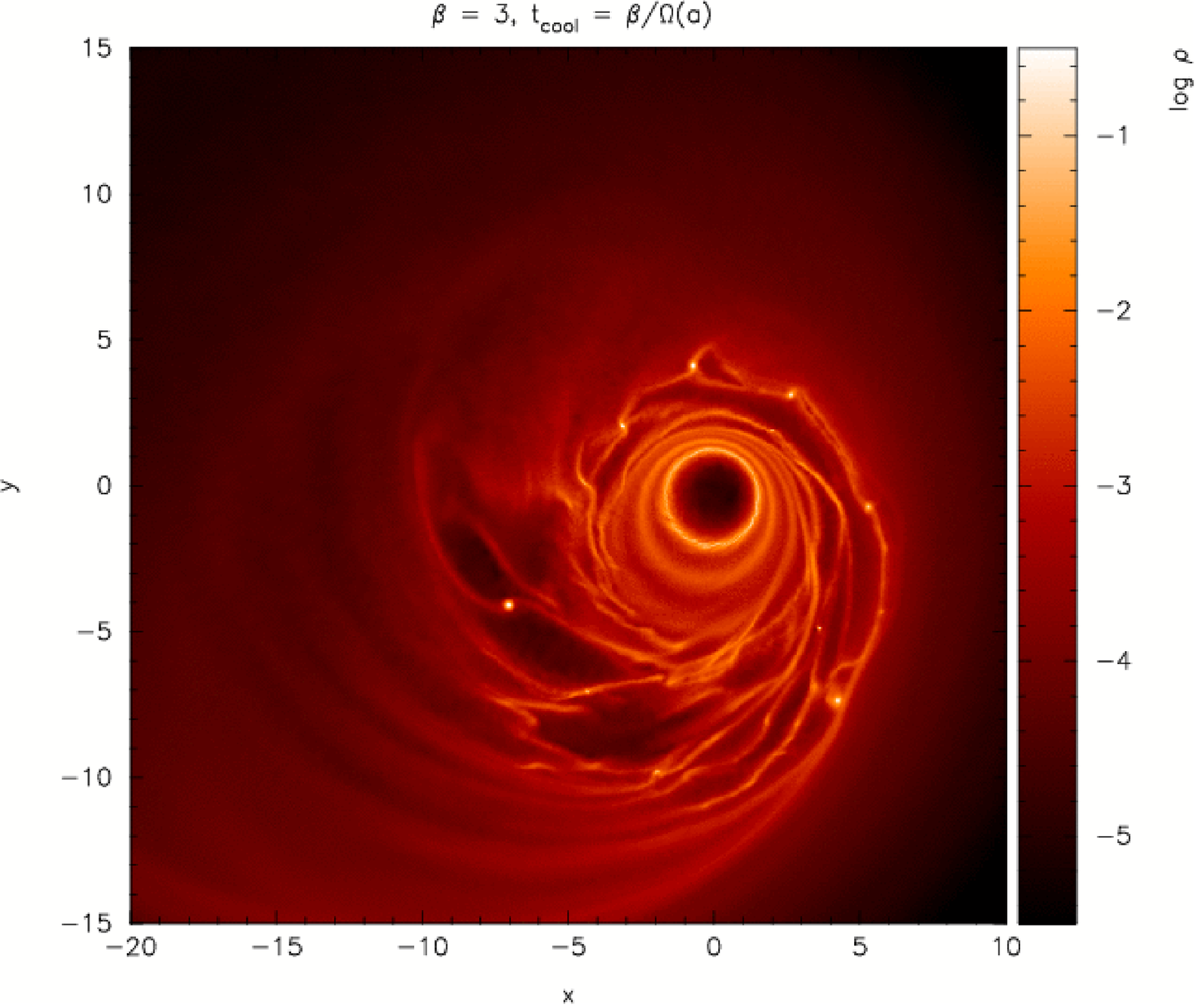}

\vspace*{3mm}

\includegraphics[angle=270,width=\hsize]{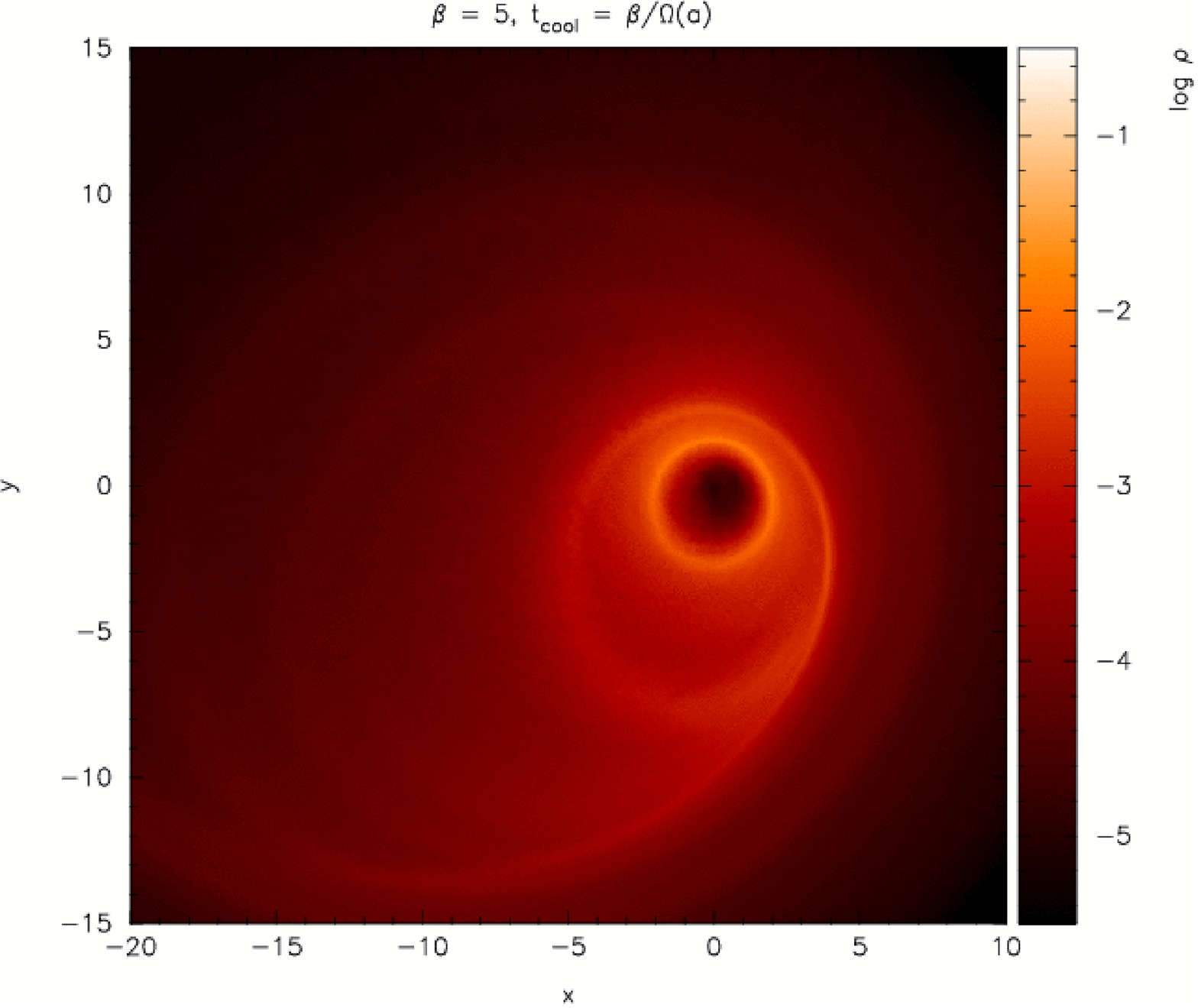}
\caption{Snapshots of disk midplane density in the central region of the low-resolution runs with $e=0.5$: the upper panel shows a model with $\beta=3$ (simulation {\sc ecc.50}), the lower $\beta=5$ (simulation {\sc ecc.50b5}).  Both are plotted on the same (logarithmic) density scale, and both are shown for $t=112.5$.  The $\beta=3$ run has fragmented, while the $\beta=5$ run remains stable.  The major axis of the disk was initially aligned with the $x$-axis: the precession of the inner disk is clearly seen in both snapshots.}\label{fig:b3_b5}
\end{figure}
Our first aim was to study the effect of eccentricity on the overall stability of the disk against fragmentation.  The energy liberated by circularization of an eccentric orbit (with a fixed angular momentum) is given simply by the product of $e^2$ and the total energy of the corresponding circular orbit, so in principle this represents a large energy reservoir that could stabilize the disk against fragmentation.  However, in our simulations this energy is not liberated sufficiently quickly to prevent fragmentation: all of the eccentric disks with $\beta=3$ fragmented (apart from that with $e=0.75$, see below), while none of the models with $\beta=5$ showed any evidence for fragmentation despite running for several outer orbital times (see Fig.\ref{fig:b3_b5})\footnote{Visualizations of our SPH simulations were created using {\sc splash}: see \citet{price07} for details.}.  Moreover, the fact that the disk with $e=0.75$ and $\beta=3$ did not fragment is likely an artefact of our numerical set-up.  In this simulation the major-to-minor axis ratio of the disk is so large that when the inner disk precesses it approaches the outer edge of the disk.  At this point density waves are reflected back off the outer edge of the disk, and appear to stabilize the disk against fragmentation.  In principle this could be a physical effect, but it is not clear that a real disk would have such a sharp outer edge with such a large major-to-minor axis ratio.  Additional simulations with a much larger dynamical range in semi-major axis (and therefore a much larger particle number) are required to investigate this further.  We therefore conclude that the disk eccentricity has little or no effect on the location of the fragmentation boundary, except possibly at very high eccentricity ($e\gtrsim 0.75$).

In a sense, this can be attributed to our choice of initial conditions.  Our disks were set up so that the angle of pericenter was constant with radius, and consequently none of the initial streamlines intersect.  We expect the circularization energy to be liberated by shocks, which arise when differential precession of the disk causes different orbital streamlines to intersect.  In our models the central body provides a point-mass potential, and the only non-point-mass contribution to the potential (i.e.,~the part that causes precession) comes from the disk itself.  The precession timescale is approximately $(M_{\mathrm {bh}}/M_{\mathrm d}) t_{\mathrm {orb}}$, but the disk becomes unstable to fragmentation after approximately 1--2 local cooling times.  Consequently our disks become unstable to fragmentation before they are able to liberate a significant fraction of their circularization energy (as seen in Fig.\ref{fig:b3_b5}), and the eccentricity has no significant effect on the overall stability criteria.  We find that a disk with uniform eccentricity is just as likely to be unstable to fragmentation as a circular disk with the same thermodynamic properties, but note that our disks are artificially constructed to have such a uniform eccentricity.  The question of whether or not these initial conditions are realistic is discussed in Section \ref{sec:dis_ics}.

\subsubsection{The effects of disk eccentricity on the fragmentation process}
\begin{figure}
\includegraphics[angle=270,width=\hsize]{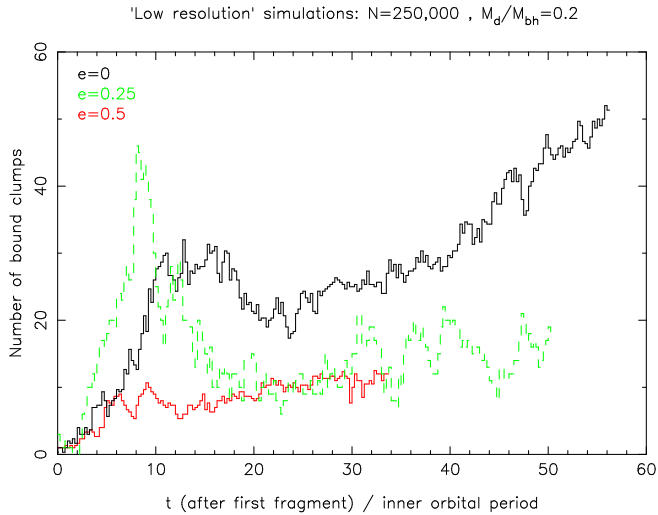}
\caption{Evolution of the number of bound clumps in simulations with different eccentricity: the time axis is normalised to the point where the first bound clump forms in each simulation.  The solid black ($e=0$) and red ($e=0.5$) lines show the mean of three different simulations; the (noisier) dashed green line ($e=0.25$) is for a single simulation only.  The eccentric disks clearly form fewer bound clumps than are seen in the circular case.  The decline in clump number at $t\simeq15$ in the circular disks is due to mergers.}\label{fig:number_low}
\end{figure}

\begin{figure}
\includegraphics[angle=270,width=\hsize]{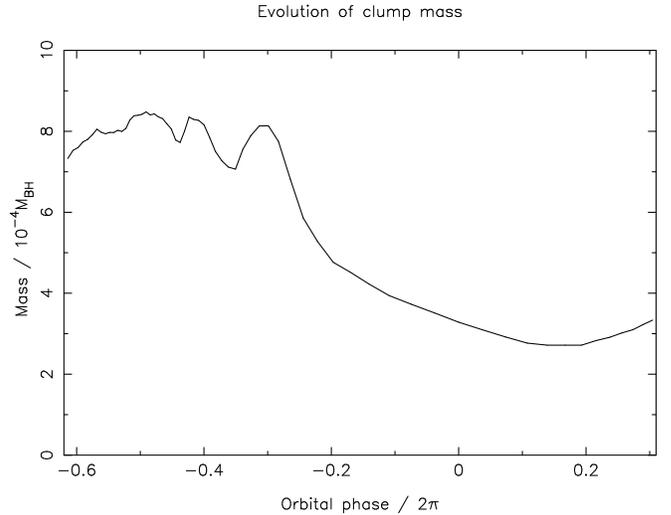}
\caption{Evolution of the mass of a typical clump, formed in a disk with $e=0.5$ (simulation {\sc ecc.50}).  The horizontal axis shows the orbital phase, with the zero-point taken at pericenter; the vertical axis shows the mass bound to the clump, in units of the mass of the central black hole.  The plot begins at the point where the clump first becomes bound, and ends where the simulation stops (giving a total period of approximately one orbit for this particular clump).  The decrease in mass caused by tidal stripping during pericenter passage is clearly visible.  The small variations in mass close to apocenter arise when the clump passes through local enhancements in the gas density, caused by global spiral density waves.}\label{fig:clump39}
\end{figure}

While it seems that eccentricity has little or no effect on the overall stability of a self-gravitating disk, our simulations do show that the details of the fragmentation process are strongly affected by eccentricity.  This is clearly seen in Fig.\ref{fig:number_low}, which shows the number of bound clumps as a function of time for simulations with different eccentricity.  Although all of the disks with $\beta=3$ fragment, clump formation is clearly suppressed in the cases where $e\ne0$, suggesting that disk eccentricity has a significant influence on the accretion of gas on to individual clumps.  This is presumably due to the variable tidal forces around an eccentric orbit: gas which is bound to a particular clump at apocenter is not necessarily bound at pericenter.  In the simulations with $e\ne0$, the most weakly-bound clumps do not survive their first pericenter passage, and instead are sheared apart by tidal forces.  The simulation with a variable cooling time (simulation {\sc ecc.50var}, not shown in Fig.\ref{fig:number_low}) shows less pronounced tidal stripping than those with constant cooling time (as the cooling is faster near pericenter), but still results in many fewer fragments than seen in the circular disks.

The effect of tidal forces is highlighted further in Fig.\ref{fig:clump39}, which shows the mass of a particular clump in one of the $e=0.5$ runs during the first orbit following its formation.  The mass of the clump is approximately constant around much of the orbit, but decreases by around 70\% as the clump passes pericenter.  This clearly demonstrates that the variable tidal forces in an eccentric disk have a strong effect on the accretion process, and that tidal stripping of this kind must be taken into account when modelling this process.

\subsection{High-resolution runs}
In order to study the effect of eccentricity on the fragmentation process in more detail, we conducted two additional simulations with much higher numerical resolution.  As described in Section \ref{sec:sim_dets}, these simulations used 5 million SPH particles and $M_{\mathrm d}/M_{\mathrm {bh}}=0.05$, resulting in an SPH particle mass of $10^{-8}M_{\mathrm {bh}}$.  In addition, we imposed a minimum SPH smoothing length on these simulations \citep{monaghan92,nayak07}, limiting collapse at the highest densities and allowing us to follow the simulations for much longer after the formation of the first fragments.  The thinner disk used here results in many more fragments being formed, and following the simulations for longer allows us to construct MFs of the formed clumps with good number statistics.  We note that the recent work of \citet{pd07} found, using an isothermal equation of state, that the artificial viscosity used in SPH can increase the survival times of weakly-bound clumps in simulations such as these.  In principle this could influence our results, so we adopt a resolution limit which ensures that all our clumps are well-resolved, and this should minimize the effects of artificial dissipation on our derived mass functions.  We use the algorithm described in Section \ref{sec:res_low} above to identify clumps, but reject clumps with fewer than 192 bound SPH particles (i.e.,~$3N_{\mathrm {ngb}}$) from our subsequent analysis as such clumps are only marginally resolved.  Scaled to the GC system, this corresponds to a resolution limit of $\simeq6$M$_{\odot}$ in mass and $\simeq20$AU in length (assuming a black hole mass of $3\times10^6$M$_{\odot}$ and a characteristic disk radius of 0.1pc).  We compute the number of bound clumps for every output snapshot (every 0.05 time units), but for computational reasons we only evaluate the MF for a handful of these snapshots (every 1 time unit, starting from the point where the first bound clump forms).  

\subsubsection{Circular disk}
\begin{figure}
\includegraphics[angle=270,width=\hsize]{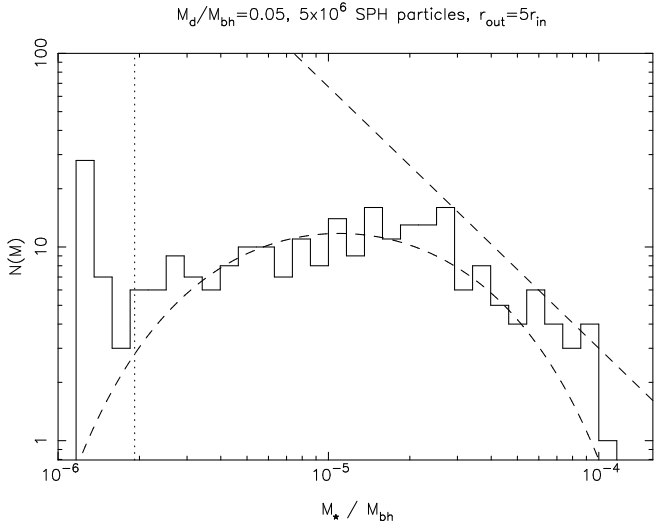}
\caption{Final MF for the circular disk, measured at the point where the simulation stopped ($t=12.7$).  The solid histogram shows the measured MF; the dashed lines show the Salpeter slope and best-fitting log-normal function (with a width of 0.79dex).  The vertical dotted line denotes the resolution limit.}\label{fig:MF_circ}
\end{figure}
The reference simulation of a circular disk ({\sc circhr}), ran for 12.7 inner orbital periods.  The first bound fragment was formed at $t=6.2$, and most of the fragmentation occurred at radii $<2$ (simply because the dynamical time in the outer disk is too long for fragmentation to occur).  At the point where the calculation stopped it had formed 313 bound clumps, of which 263 were above the (mass) resolution limit.  Within $r=2$ the bound clumps account for 40.5\% of the total disk mass, suggesting that much of the gas has been accreted into bound objects and the resulting MF is indeed representative.  The final MF is shown in Fig.\ref{fig:MF_circ}.  The slope is close to Salpeter at high masses, but the MF shows a well-resolved turnover at $\simeq 10^{-5}M_{\mathrm {bh}}$ and is in fact well-fit by a \citet{ms79} log-normal distribution function, with a width of $\simeq 0.8$dex and a characteristic mass of $1.1\times10^{-5}M_{\mathrm {bh}}$.  Note that this is essentially the Toomre mass, which for this simulation is $\pi \Sigma H^2 \simeq 10^{-5}M_{\mathrm {bh}}$: this corresponds to a characteristic mass of $\simeq 30$M$_{\odot}$ when scaled to the GC.  The MF is clearly not consistent with the Salpeter slope at even moderate stellar masses, and the low-mass turnover is consistent with previous numerical simulations \citep{nayak07}, and also with observations of the GC stellar disks \citep{ns05,paumard06}.  Thus we are confident that our reference model is both numerically accurate and physically plausible, and now look to the differences that result when the disk is eccentric.

\subsubsection{Eccentric disk}
\begin{figure}
\includegraphics[angle=270,width=\hsize]{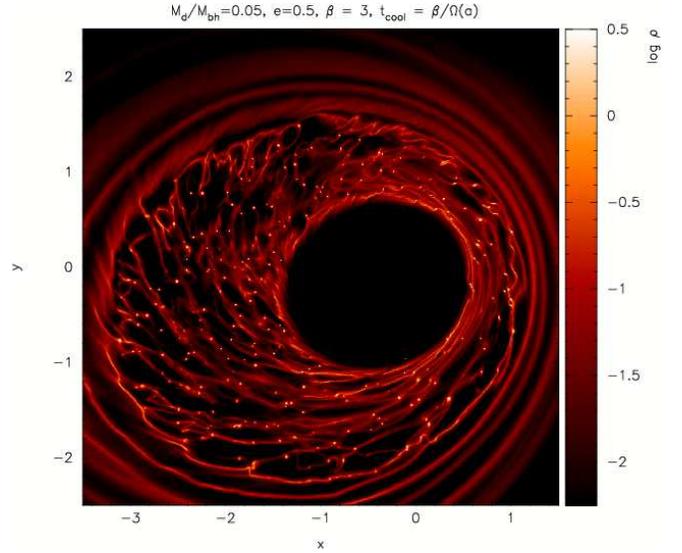}
\caption{Snapshots of disk midplane density in the central region of the high-resolution run with $e=0.5$ (simulation {\sc ecchr}).  The snapshot is taken at the point where the simulation ended, $t=13.5$, and many bound clumps are clearly visible.  The major axis of the disk was again initially aligned with the $x$-axis: the less massive disk results in considerably less precession than was seen in the low-resolution runs.}\label{fig:HR_snap}
\end{figure}
The simulation {\sc ecchr} ran for 13.5 inner orbital periods, and the first fragment formed at $t=6.75$.  As in the circular disk, the fragmentation was confined to $a\lesssim2$, and both simulations ran for $\simeq7$ inner orbital periods after the formation of the first fragment.  As we saw in the low resolution runs, the eccentric disk formed fewer clumps: at the point where this calculation stopped it had formed 189 clumps, of which 178 had masses above the resolution limit.  Within $a=2$ the bound clumps had accreted 26.5\% of the available gas.  The clumps follow orbits very similar to the disk: while there is some scatter in the instantaneous values of the clump eccentricities, the rms eccentricity of all of the bound clumps at the end of the simulation was $\simeq 0.49$ (essentially equal to the disk eccentricity of 0.5).  The final midplane density profile is shown in Fig.\ref{fig:HR_snap}.

\subsubsection{Mass function evolution}
\begin{figure}
\includegraphics[angle=270,width=\hsize]{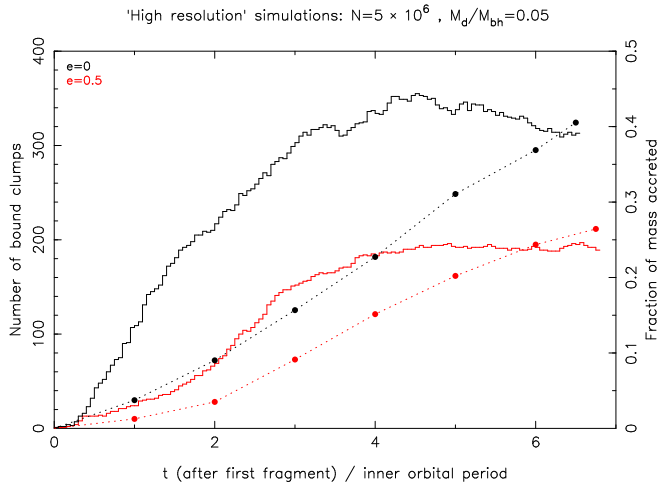}
\caption{Evolution of bound clumps in the high-resolution simulations.  The solid lines show the evolution of the number of bound clumps (left-hand scale), while the points show the fraction of disk mass within $a\le2$ accreted into bound clumps (right-hand scale).  The dotted lines between the points are added as a guide to the eye.  As in Fig.\ref{fig:number_low}, the time axis is normalized to the point where the first bound clump forms in each simulation.}\label{fig:number_high}
\end{figure}

In order to look at the effects of eccentricity on the fragmentation and accretion processes, it is instructive to look at the evolution of the MF in both the circular and eccentric disks.  Fig.\ref{fig:number_high} shows the evolution of both clump number and accreted mass in the inner part ($a\le2$) of both the circular and eccentric disks.  Both disks show a ``two-phase'' formation process: a fragmentation phase, where the clump number increases rapidly and most of the bound mass is in small clumps; followed by an accretion phase, where the clump number remains approximately constant but the clumps continue to accrete gas from their surroundings.  Once again we see that the eccentric disk forms fewer clumps than the circular disk, and we again attribute this to the tidal destruction of weakly-bound clumps in the eccentric disk.  Moreover, the accretion of mass is much slower in the eccentric disk, suggesting that tidal forces affect the accretion process across the full range of clump masses.  This is reinforced by the measured MFs: the eccentric disk MF consistently shows a larger characteristic clump mass than seen the circular disk.  

\begin{figure}
\includegraphics[angle=270,width=\hsize]{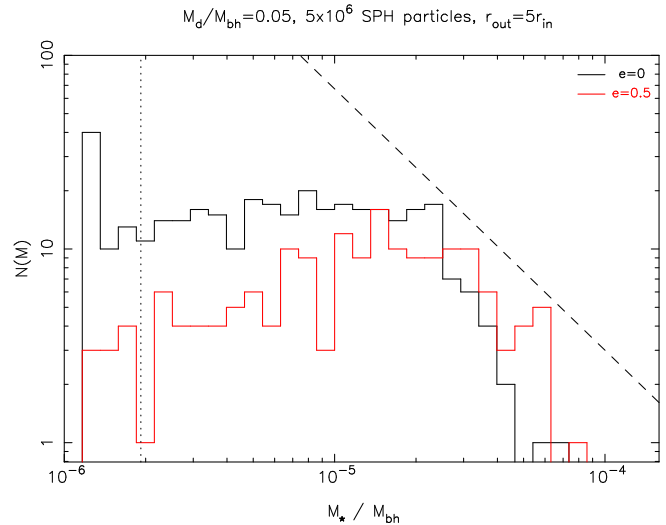}
\caption{Comparison of clump MFs in the circular (black) and eccentric (red) disks.  As in Fig.\ref{fig:MF_circ}, the dotted line shows the resolution limit, and the dashed line the Salpeter slope.  The MF for the eccentric disk is measured at the end of the simulation; in the circular case it is measured where the same fraction of the disk mass has been accreted into bound clumps.  The eccentric disk clearly forms fewer low-mass objects than the circular disk, and a KS test rejects the possibility that the two MFs are drawn from the same underlying distribution.}\label{fig:MF_comp}
\end{figure}

The MFs are compared directly in Fig.\ref{fig:MF_comp}, measured at a point where both simulations have accreted the same fraction of gas into bound clumps.  The eccentric disk MF is taken at the end of the simulation (when 26.5\% of the disk mass has been accreted into clumps), and the ``reference'' circular disk MF is taken 4.4 time units after the formation of the first clump (26.8\% accreted).  Both are reasonably well-fit by log-normal distribution functions: the best-fitting log-normal in the circular case has a characteristic mass of $6.6\times10^{-6}$ and a width of 0.6 dex; in the eccentric case the best-fitting characteristic mass is $1.2\times10^{-5}$, and the width is 0.5 dex.  A KS test gives a probability of $7\times10^{-6}$ that the two MFs are drawn from the same underlying distribution, and similar comparisons at earlier times in the simulations (with the same accreted mass fractions) all result in probabilities of $<1$\% that the two MFs are the same.  Thus we conclude that the disk eccentricity alters the accretion process significantly, primarily because of tidal effects around the eccentric orbit.  Mass is tidally stripped from bound clumps during pericenter passage, with two notable effects.  Firstly, the most weakly-bound, lowest-mass clumps fail to survive pericenter passage, resulting in a relative dearth of low-mass objects in the clump MF.  Secondly, the overall accretion rate is slowed, suggesting that an eccentric disk will take significantly longer than a circular one to accrete all of its mass into bound clumps.

\section{Discussion}\label{sec:dis}

%%%%%%%%%%%%%%%%%%%%%%%%%%

%\section{Discussion}\label{sec:dis}
While our simulations are scale-free, they have obvious applications to the formation of the stellar disk(s) observed at the GC.  Observations of the GC stellar disks find that the stellar MF must be much more ``top-heavy'' than the standard Salpeter MF.  The relatively low X-ray flux from the GC region suggests a significant deficit of low-mass ($\lesssim 5$M$_{\odot}$) stars compared to local star-forming regions \citep{ns05}, and estimates of the stellar MF from the $K$-band luminosity function suggests a MF slope that is 1.0--1.5 dex flatter than the Salpeter slope \citep{paumard06}.  Also, given that the total stellar mass in the disk(s) is $\sim10^4$--$10^5$M$_{\odot}$ \citep[e.g.,][]{ndcg06,paumard06}, the large number ($\gtrsim 100$) of observed massive O- \& WR-stars suggests a characteristic stellar mass of $\simeq 10$--50M$_{\odot}$, much larger than that observed in the solar neighbourhood.  This is consistent with our results \citep[and those of previous studies such as][]{nayak07}, which predict a characteristic stellar mass of $\simeq 30$M$_{\odot}$ (when scaled to the GC), and a MF that is significantly deficient in low-mass objects.  Previous studies have not been able to reproduce all of the observed properties of the disks (notably the eccentric stellar orbits), but our results suggest that fragmentation of moderately eccentric accretion disk can form disks of stars with masses and orbits consistent with those observed.  Our simulations suggest that eccentric disks fragment to produce MFs that are slightly more top-heavy than expected in the circular case (due to the tidal destruction of the lowest-mass clumps), but not so top-heavy as to be inconsistent with the observed stellar populations.  Moreover, previous simulations have shown that changes in the thermodynamic properties of the disk, such as the cooling rate, can alter the MF at a level at least as significant as that seen in our simulations \citep{nayak07}.  Consequently, we are satisfied that our results are robust, within the range of parameter space explored by our simulations.  However, there are three major areas in which our simulations differ from real systems, and we discuss each of these in turn below.

In addition, we note that angular momentum transport (by spiral waves) is negligible in the simulations that exhibit fragmentation, as the fragmentation and subsequent accretion of gas occurs on the dynamical timescale (which is much shorter than the timescale for angular momentum transport).  It has long been thought that angular momentum transport due to gravitational instabilities can be responsible for significant accretion in disks around black holes \citep[e.g.,][]{sfb89}.  This is indeed likely, but our results confirm those of \citet{nayak07}, that star formation due to gravitational instability in accretion disks occurs on a timescale that is significantly shorter than the timescale for angular momentum transport.  Consequently, if star formation occurs at all radii in the disk, it seems likely that little or no gas will be accreted on to the central black hole.  We note, however, that real disks around black holes are likely only unstable to fragmentation in their outer regions \citep[e.g.,][]{levin07,kp07}, and that whether or not gravitational instability in such disks results in fragmentation or transport depends critically on the issues of cooling discussed in Section \ref{sec:dis_cool}.

Our simulations are most applicable to the GC stellar disk(s), but their scale-free nature allows us to consider the results in terms of other systems also.  It has been suggested that the young stellar population at the GC may be characteristic of more generic episodes of black hole (AGN) activity in other galaxies \citep[e.g.,][]{lb03,tqm05}, and this has implications for models of AGN accretion and the growth of supermassive black holes.  The consequences of our models for AGN feeding depend on the disk thermodynamics (see Section \ref{sec:dis_cool}), but we note in passing that our models fit naturally into, for example, the black hole growth scenario presented by \citet{kp06,kp07}.  The applicability of our results to protoplanetary discs, however, is less clear.  It seems likely that protoplanetary disks are gravitationally unstable at early times in their evolution \citep[e.g.,][]{durisen_ppv}, but it is less clear that such disks possess significant eccentricities.  If eccentric, self-gravitating disks do exist around young stars then our results could easily be applied to such systems, but at present there is no strong evidence for the existence of such systems.

\subsection{Initial configuration}\label{sec:dis_ics}
The first area of concern is in our initial conditions.  We choose an initial configuration that is marginally stable, and allow the disks to cool into instability.  This approach has been followed by previous studies \citep[e.g.,][]{rice03,rice05}, and recent work suggests that use of these artificial initial conditions does not affect the fragmentation boundary in circular disks \citep{clarke07}.  What is less clear, however, is whether or not our choice of a uniform angles of pericenter is similarly valid.  As discussed in Section \ref{sec:intro}, in systems like the GC we expect the disk to form via some type of accretion event, and while previous simulations \citep[e.g.,][]{sanders98} have found that an eccentric configuration can result, there is no particular reason to expect a uniform eccentricity.  If the disk does not have uniform eccentricity then neighboring orbital streamlines will intersect, and the resulting shocks have the potential to provide significant additional heating.  Such heating may act to stabilize the disk against fragmentation, but a realistic treatment of this problem requires full (magneto-)hydrodynamic simulations of the infall and capture of a gas cloud by the black hole, and such modeling is clearly beyond the scope of this work.  What we have demonstrated, however, is that if an eccentric disk is able to evolve towards a configuration where it satisfies the (circular) conditions for fragmentation, then it will fragment before any such shock heating is able to stabilize the disk.

\subsection{Cooling and thermodynamics}\label{sec:dis_cool}
The second major area of uncertainty in our models is our simplified cooling prescription.  Our cooling prescription (Section \ref{sec:cooling}) parametrizes the cooling rate by semi-major axis only, and while this is ideal for the sort of numerical experiments conducted here, it is far from realistic.  In the case of protoplanetary disks, the subject of heating and cooling rates in gravitationally unstable disks is an area of on-going debate \citep[e.g.,~the review by][]{durisen_ppv}, but in the case of a disk around a black hole we can justify the approximate form of our cooling prescription by a few simple scaling arguments.  Similar arguments have been discussed previously by other authors \citep[e.g.,][]{lb03,jg03,gt04,levin07,rafikov07}, but we re-state them here for completeness.

A self-gravitating disk has $Q\simeq 1$, and therefore if we know the disk surface density we can estimate the sound speed, and therefore the temperature, in the disk.  If we scale our model to the GC system (by adopting a characteristic radius of 0.1pc and $M_{\mathrm {bh}}=3\times10^6$M$_{\odot}$), we find a typical disk surface density of $\sim 100$g cm$^{-2}$ and a corresponding disk temperature of $\sim 100$K (estimated at $a=2$, but these quantities do not vary significantly over the limited radial range of our simulations).  At such temperatures the dominant source of opacity is dust grains, and typical values for the Rosseland mean opacity are $\kappa \simeq 1$--10cm$^2$g$^{-1}$ \citep[e.g.,][]{semenov03}.  The opacity does not change dramatically at temperatures below that at which the dust grains sublime (typically $\simeq 1500$K).  As seen above, scaling our models to the GC system (assuming that the initial gas mass in the disk is comparable to the observed stellar mass) results in a much lower temperature than this, so significant variations in the disk opacity (such as the opacity gap discussed by, e.g., \citealt{jg03,tqm05}) seem unlikely.  Only if star formation in such disks is very inefficient, contrary to what is seen in our (and other) simulations, would we expect the disk temperatures to be high enough for dust sublimation to be significant.  Therefore, we expect such a disk to be marginally optically thick\footnote{Note that this also verifies our earlier assumption that the cooling rate is unlikely to change significantly around the orbit.}, and the resulting cooling timescale is typically comparable to the orbital period \citep[e.g.,][]{rafikov07}.  However, we also note that our cooling model takes no account of changes in the cooling function as the local density increases (due to gravitational collapse).  A more realistic model would make use of an opacity-based cooling model that accounts for the density structure of the disc, and would also adopt a somewhat stiffer equation of state in the the collapsing clumps.  The application of such models to protoplanetary disks is still mired in controversy \citep[e.g.,][]{boley06,mayer07}, but the physical conditions in a GC disk are known to be much more conducive to fragmentation than those in protoplanetary disks \citep[see e.g.,~the discussion in][]{rafikov07}.  Consequently we expect that our disks will cool rapidly enough to be unstable to fragmentation.  Moreover, the uncertainties discussed here are not specific to our models of eccentric disks, so we do not expect our results regarding the differences between eccentric and circular disks to be affected by these simplifications.

A further simplification is that our cooling prescription depends only on the local conditions in the disk, and is not influenced at all by the global disk structure.  Previous work using this scale-free, local cooling prescription has found that it gives rise to instabilities that are well-approximated by a local description \citep{lr04,lr05}, and therefore the global disk structure is unlikely to have a significant effect on the spatial distribution of clumps in our simulations.  However, in a real disk the cooling time is not scale-free and, in general, gravitational instabilities cannot be well-described locally \citep{bp99}.  Numerical simulations of protoplanetary disks have found that transport can be dominated by low-order global modes in the non-scale-free case \citep{boley06,cai07,durisen_ppv}.  It seems likely, however, that the local approximation will be more accurate for the very thin disks considered here than in the case of  protoplanetary disks (where $M_{\mathrm d}/M_* \sim 0.1$).

What is less certain is whether or not a realistic system is able to evolve into the configuration discussed here.  The virial temperature of a gas cloud that falls towards the GC from, for example, the circumnuclear ring is orders of magnitude larger than the disk temperatures considered here \citep[e.g][]{morris93,sanders98}.  An infalling cloud may not virialize before it forms a coherent disk structure, but there is clearly sufficient energy available to heat the gas significantly.  At temperatures from $\simeq 1000$--$10^4$K the typical Rosseland opacities are much smaller than those at $\sim 100$K, as the primary coolants (dust, and also molecular species) are destroyed at such temperatures \citep{semenov03}.  Thus, we expect the cooling rates at higher temperatures to be much lower than those considered here, and it may be difficult for such a disk to cool into instability in a more realistic model \citep[see also the discussion in][]{jg03}.  Our simple thermodynamic model is viable once the disk reaches the state considered here but, as discussed above, we do not address the issue of whether or not a real system will be able to reach such a configuration.

\subsection{Accretion physics at stellar scales}
The last major area neglected by our simulations is the small-scale physics of star-formation.  Our simulations have sufficient numerical resolution to treat the fragmentation process correctly, but when scaled to the GC our minimum length resolution is a few tens of AU.  This is 2--3 orders or magnitude larger than the typical stellar radius, and we make no attempt to model the physics of star formation on small scales.  Essentially our MFs are MFs of pre-stellar cores, and as such they may not be representative of the stellar MFs that result from disk fragmentation.  However, we note that the fragmentation into bound clumps occurs on orbital timescales, and the results presented in Fig.\ref{fig:number_high} suggest that all of the disk gas will be bound to clumps within $\simeq10$ orbital periods.  The orbital period of the GC stellar disks is $\sim 1000$yr, but the typical timescales for such cores to collapse to form stars are $\sim 10^4$--$10^5$yr \citep[e.g.,][]{mt03}.  Consequently we do not expect stellar feedback processes to have a significant effect on the process of disk fragmentation.  However, individual bound clumps may well form more than one star each, and we also expect to see some mergers between clumps \citep{levin07}.  More detailed simulations, using higher numerical resolution and more realistic thermal physics, will be required to address these issues further. 

%%%%%%%%%%%%%%%%%%%%%%%%%%%%

\section{Summary}\label{sec:summary}
We have constructed hydrodynamic models of gravitationally unstable disks, and investigated the effects of eccentricity on disk fragmentation.  We find that the fragmentation boundary is unaffected by eccentricity, and that conditions which lead to fragmentation in circular disks do likewise in eccentric disks.  However, we find that the fragmentation process is altered by eccentricity, as the variable tidal forces around an eccentric orbit have a strong effect on the accretion of gas on to bound fragments.  We find that the formation of low-mass fragments is suppressed, and that the growth (via accretion) of more massive fragments proceeds much more slowly in an eccentric disk than in a circular one.  We consider our results in the context of the GC stellar disks, and find good agreement between the MFs in our simulations and those observed.  We find that fragmentation of eccentric accretion disks is a viable mechanism for the formation of the GC stellar disks, with the resulting stellar MF and orbits consistent with observations.

%%%%%%%%%%%%%%%%%%%%%%%%%%%%
\acknowledgments
We thank Volker Springel for providing us with some subroutines that are not part of the public version of {\sc gadget2}, and thank an anonymous referee for several valuable comments.  This work made use of NSF (NCSA) super-computing allocations AST070006N and AST070011, and was supported by NASA under grants NAG5-13207, NNG04GL01G and NNG05GI92G from the Origins of Solar Systems, Astrophysics Theory, and Beyond Einstein Foundation Science Programs, and by the NSF under grants AST--0307502 and AST--0407040.  

%%%%%%%%%%%%%%%%%%%%%%%%%%%%

\end{document}